\documentclass
[prb,superscriptaddress,superbib,endfloats,nofootinbib,nobibnotes,titlepage,byrevtex,amsfonts,amssymb,onecolumn,preprint,final,showkeys]{revtex4}%
\usepackage{amsfonts}
\usepackage{amsmath}
\usepackage{amssymb}
\usepackage{graphicx}
\usepackage{caption}%
\providecommand{\U}[1]{\protect\rule{.1in}{.1in}}

\begin{document}
\title{Changes of Interatomic Force Constants Caused by Quantum Confinement Effects:
Study on the Calculations for the First-order Raman Spectrum of Si
Nanocrystals in Comparison with Experiments.}
\author{Wei-Shan Lee}
\email{weishan.lei@gmail.com}
\affiliation{Department of Physics, National Taiwan University, Taipei 115, Taiwan,
Republic of China.}
\affiliation{Nano Science and Technology Program, Taiwan International Graduate Program,
Academia Sinica, Taipei 115, Taiwan, Republic of China.}

\begin{abstract}
The redshifts and asymmetric broadening observed in nanocrystal Raman Spectra
are attributed to the quantum confinement effects by some authors. But others
show that they may come from the local heating caused by the incident laser as
well. In this study we demonstrate that in the Si nanocrystal case the latter
at most has obvious effects on the broadening but has negligible effects on
the 1LO peak shift, while the former contributes most of the 1LO peak shift.
We also demonstrate that after assigning appropriate interatomic force
constants in the calculation of Raman Spectrum by bond polarizability
approximation model within the regime of free boundary condition, we may
acquire the matching 1LO peak shift with experiments.

\end{abstract}
\maketitle

\section{\bigskip\bigskip\bigskip Introduction}

Attributed to its compatibility to the state-of-the-art semiconductor
manufacturing process and the abundance of silicon in nature, silicon
nanocrystals, or in a more favorable name, quantum dots, raise world-wide
interests for decades in studying their quantum confinement effects on
electronic and optical properties\cite{PRB 47 1397 1993}$^{\text{-}}%
$\cite{laser photonic review} because of the promising applications to devices
such as the light emitting diodes\cite{J Lum 80 263 1999} and single electron
transistors\cite{Superlattices and Microstructures 28 177 2000}. In addition,
the vibrational properties of the Si nanocrystals also attract great
attention\cite{JAP 78 6705 1995}$^{\text{-}}$\cite{PRB 80 193410 2009}. For
one thing, Raman spectroscopy is a noninvasive technique to investigating
defects or qualities of devices. For example, the first order Raman shifts may
be used to examine the crystalline quality of the silicon
nanostructures\cite{nanotech 18 175705 2007}$^{\text{,}}$\cite{JAP 90 4175
2001}. On the other hand, studies on electron-phonon coupling and
thermodynamic properties of silicon nanocrystals require full understandings
for the behaviors of phonon confinements. In comparisons with the first order
Raman shifts, the second order Raman shifts may be enhanced by
electrochemically etched silicon substrates that are immersed in different
duration of time\cite{Chem Phys Letter 382 502 2003}. Likewise, Si
nanocrystals made by annealing hydrogenated amorphous silicon($\alpha$-Si:H)
with continuous-wave laser are also used to study the first and the second
Raman spectra\cite{PRB 64 073304}.

In study of the first-order Raman Spectrum of Silicon nanocrystals, there are
at least two prominent phenomena to which we should pay attention. First, the
asymmetric broadening towards low frequency around 1LO transition (520
cm$^{\text{-1}}$for bulk) is observed in the spectrum, indicating a phonon
confinement effect. Secondly, the (bulk) 520 cm$^{\text{-1}}$ peak shift to
lower frequency with reducing nanocrystal sizes. There have been extensive
theoretical and experimental studies of the first-order nanocrystal Raman
scattering. H.Richter et.al.\cite{Richter's}\textit{, }describes a
phenomenological exponential function that restricts the nanocrystal phonon
wavefunctions in the sphere. Several different weighting functions are also
used\cite{SSC Campbell and Fauchet}$^{\text{-}}$\cite{PRB 73 033307 2006}.
However, even though it is claimed that experimental data could be well
explained by the above method, it is still obscure for the exponential and
weighting functions to have definitive physical meanings so that they may not
provide information of interatomic physical quantities. Therefore, it could be
more preferable that we start from solving the characteristic vibrations of
the nanocrystal, which contribute to the change of polarizability resulting in
the Raman effects. Thus, the bond polarizability approximation\cite{BPA
Bell}(BPA) is introduced to calculate the first-order Raman Spectrum of
Si\cite{Cheng and Ren} as well as fullerene\cite{BPA fullerene}. This model
suggests that the total polarizability tensor is the sum over each axially
symmetric polarizability tensor connecting each pair of atoms $ij$, for which
may be expressed as a function of two bond-length dependent parameters, one
referring to mean polarizability while the other anisotropic polarizability.
The total polarizability tensor is then approximated up to the first order
with respect to the equilibrium position of the atoms. However, the asymmetric
broadening effect does not meet agreement with the experimental results in the calculation.

Failure to success may be imputed to several reasons. First, it is difficult
to obtain appropriate interatomic force constants (IFCs) among atoms by only
considering the first-neighbor interactions, resulting in unsuitable phonon
eigenfrequencies and eigenvectors. Improvements may be made by considering the
IFCs up to the second neighbors. Secondly, there seems no reason to only
consider the first-neighbor atoms in the parameters of bond polarizability.
Thirdly, some studies show that the peak shift and asymmetric broadening may
be due to local heating from the tiny focusing laser spot rather than size
effects\cite{PRB 80 193410 2009}$^{\text{,}}$\cite{PRB 66 161311R 2002 local
heating}. However, since both of the parameters of bond polarizability and the
IFCs cannot be known \textit{in priori}, it is difficult to obtain proper
values of them without comparing the peak shift of experiments. Therefore,
close agreement of asymmetric broadening in concomitant of the peak shift
towards low frequencies between calculations and experiments remains challenging.

In this study, we first obtain IFCs by fitting the bulk silicon dispersion
relations calculated by the rigid-ion model\cite{RIM} up to the second nearest
neighbors with the experimental data from the neutron scattering at the three
special points, $\Gamma$, $X$, and $L$, in the First Brillouin zone. The IFCs
are then compared with the results from ab initio calculations in order to
show the significance of the parameters. Afterwards, these fitting parameters
are used to calculate the vibrational properties in nanocrystals by little
modification with the self energy of the surface atoms. Nanocrystals with two
different diameters are made and measured by Raman spectrometer with He-Ne and
Argon-ion lasers. Peak shift and broadening effect regarding to the
nanocrystal sizes or laser power variations are discussed. The first order
Raman spectra are calculated by bond polarizability approximation
(BPA)\cite{BPA Bell}. It is found that we may acquire close peak shift by
assigning proper IFCs in the calculation and obtain more satisfactory
asymmetric broadening effects by considering the parameters of polarizability
up to the second-nearest neighbor atoms. Fair resemblance to the calculated
first-order Raman spectra with the experimental ones is seen.

\section{Theory and Sample Preparations}

\subsection{The Rigid-ion Model and Interatomic Force Constants in Bulk}

\begin{figure}
[t]
\begin{center}
\includegraphics[
trim=1.301128in 1.244918in 0.000000in 0.683656in,
height=3.8692in,
width=5.7147in
]%
{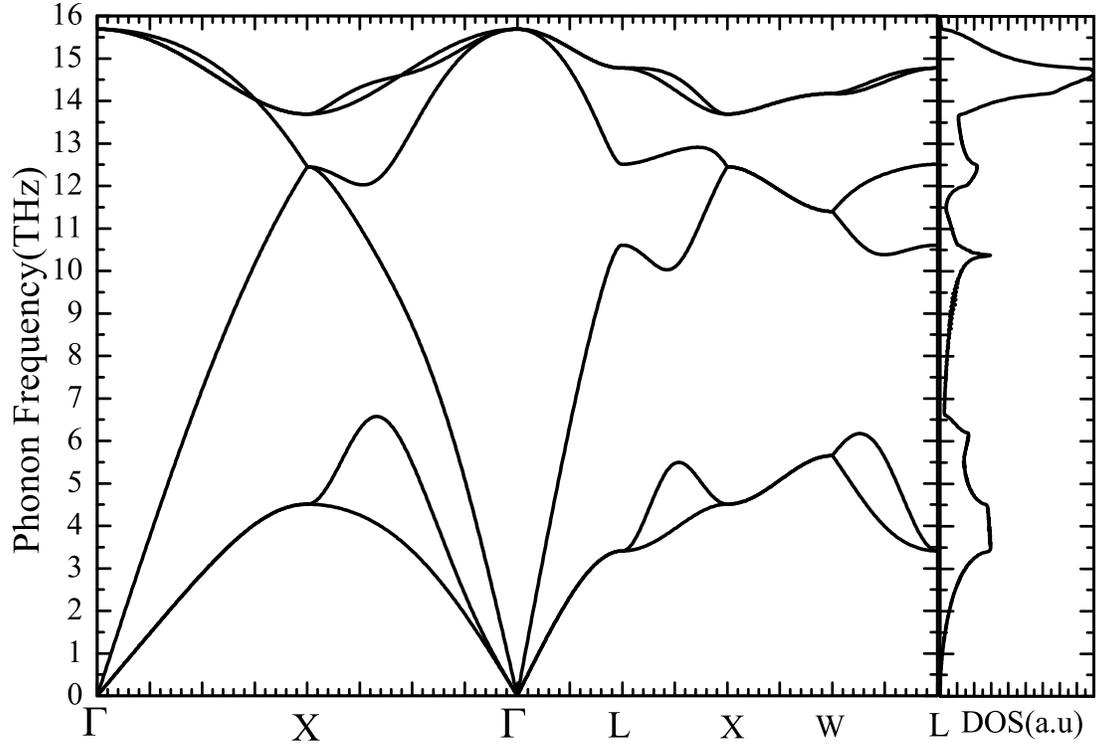}%
\caption{Phonon dispersion relation and density of states calculated by Rigid
Ion Model with the parameters obtained from solving the multiple equations at
the three special points in the first Brillouin zone as well as $E=0.0189$.}%
\label{PDR and DOS}%
\end{center}
\end{figure}

For a homopolar Silicon bulk crystal, the potential energy $U$ is expanded up
to the second order harmonic approximation with respect to the equilibrium
position of an atom. This term, $\frac{\partial^{2}U}{\partial u_{i}\partial
u_{j}}|_{\overset{\rightharpoonup}{x_{0}}}$, is then described as the dynamic
matrix $D_{jj^{\prime}}^{\sigma\sigma^{^{\prime}}}$in the eigenvalue problem,%
\begin{equation}
\underset{\sigma^{^{\prime}}j^{\prime}}{%
{\displaystyle\sum}
}D_{jj^{\prime}}^{\sigma\sigma^{^{\prime}}}(\overset{\rightharpoonup}%
{q})u_{j^{\prime}}^{\sigma^{^{\prime}}}=M_{\sigma}\omega^{2}(\overset
{\rightharpoonup}{q})u_{j}^{\sigma} \label{eval_problem}%
\end{equation}
, where $\sigma$ denotes different kinds of atoms in a unit cell and
$j^{^{\prime}}$ denotes x,y,z in Cartesian coordinates, $M_{\sigma}$ being the
mass of the atom for each kind. Furthermore, for periodicity in bulk,%
\begin{equation}
D_{jj^{\prime}}^{\sigma\sigma^{^{\prime}}}(\overset{\rightharpoonup}%
{q})=\underset{\overset{\rightharpoonup}{R}}{%
{\displaystyle\sum}
}D_{jj^{\prime}}^{\sigma\sigma^{^{\prime}}}(\overset{\rightharpoonup}%
{R}+\overset{\rightharpoonup}{S}_{\sigma}-\overset{\rightharpoonup}{S}%
_{\sigma^{\prime}})e^{i\overset{\rightharpoonup}{q}\cdot(\overset
{\rightharpoonup}{R}+\overset{\rightharpoonup}{S}_{\sigma}-\overset
{\rightharpoonup}{S}_{\sigma^{\prime}})} \label{DM_periodic_condition}%
\end{equation}
, where $\overset{\rightharpoonup}{R}$ denotes the position vectors of the
first-neighbour or second-neighbour atom and $\overset{\rightharpoonup}%
{S}_{\sigma}$denotes the vectors in which the atom in the primitive unit cell
with species $\sigma$ resides. In the rigid-ion model\cite{RIM}, the term
$D_{jj^{\prime}}^{\sigma\sigma^{^{\prime}}}(\overset{\rightharpoonup}%
{R}+\overset{\rightharpoonup}{S}_{\sigma}-\overset{\rightharpoonup}{S}%
_{\sigma^{\prime}})$ is described by several 3 by 3 force-constant matrices.
Specifically, the 3 by 3 matrix for the first-nearest neighbour atom at the
position $\frac{a}{4}(1,1,1)$ is defined by%
\begin{equation}
D^{1,2}(1)=\left(
\begin{array}
[c]{ccc}%
A & B & B\\
B & A & B\\
B & B & A
\end{array}
\right)  \label{DM_1st_nb}%
\end{equation}
, and for the second-nearest neighbour atom at the position $\frac{a}%
{4}(2,2,0)$ we define%
\begin{equation}
D^{\sigma\sigma}(1)=\left(
\begin{array}
[c]{ccc}%
C & D & E\\
D & C & E\\
-E & -E & F
\end{array}
\right)  \label{DM_2nd_nb}%
\end{equation}
. For all the other atoms at the position $\overset{\rightharpoonup}{R}$, the
relative force-constant matrices may be derived by similarity transformation
$D^{\sigma\sigma^{^{\prime}}}(\overset{\rightharpoonup}{R})=T_{i}%
D^{\sigma\sigma^{^{\prime}}}(1)T_{i}^{-1}$ with orthogonal matrix $T_{i}$ of
determinant 1\cite{book methods in computational physics}. Here only up to
second-order nearest neighbours are taken into consideration. We derive the
eigenvalues at the $\Gamma,X,L$ points in the Brillouin zone\cite{PRB 74
054302 2006}.\bigskip\ For $\overset{\rightharpoonup}{q}=\Gamma$, the
eigenvalues are \thinspace$0$ and $-8A$, both being triplet degenerate. For
$\overset{\rightharpoonup}{q}=X$ at $(1,0,0)$, we have$\bigskip$
\begin{subequations}
\begin{align}
&  -4A-16C\text{ ; one LO and one LA}\\
&  -4A-4B-8C-8F\text{ ; two TO}\\
&  -4A+4B-8C-8F\text{ ; two TA}%
\end{align}

\bigskip They are all double degenerate. At last, for the $L$ point $(\frac
{1}{2},\frac{1}{2},\frac{1}{2})$, we have%
\end{subequations}
\begin{subequations}
\begin{align}
&  -6A-2B-8C+4D-4F\text{ ; two TO}\\
&  -6A+4B-8C-8D-4F\text{ ; one LO}\\
&  -2A-4B-8C-8D-4F\text{ ; one LA}\\
&  -2A+2B-8C+4D-4F\text{ ; two TA}%
\end{align}

$A,B,C,F$ are solved unambiguously at $\Gamma$ and $X$. Unfortunately one may
immediately notice that there is no value for $D$ that may simultaneously
satisfy Eq 6. Furthermore, there is no information for $E$ at the three
special points. However, one may obtain values of $D$ and $E$ by calculating
the interatomic force constants with ab initio calculations. We use linear
response\cite{Phys Rev B 55 10337 1997}$^{\text{,}}$\cite{Phys Rev B 55 10355
1997} implantation of DFT with a local density approximation (LDA) to exchange
correlation effect and Troullier-Martins pseudopotential in Abinit
code\cite{Computer Phys Commun 180 2582 2009}$^{\text{,}}$\cite{Zeit
Kristallogr 220 558 2005}. The cutoff energy is set to 30 hatrees for the
plane-wave basis. Two sets of the parameters obtained either by directly
solving the multiple equations above, from plugging in neutron scattering
experiment data\cite{data1}$^{\text{,}}$\cite{data2}, or by the ab initio
calculations are summarized and compared in Table I. The phonon density of
states with even-grid k points\cite{even grided k points} in the first
Brillouin zone and phonon dispersion relations are shown in Fig
\ref{PDR and DOS}.

One researcher said that the parameter $E$ in the IFCs is zero because
\textquotedblleft for Group IV, we must have $E_{1}$ $=$ $E_{2}$ $=0$ to
preserve symmetry.\textquotedblright\ While in compounds such as AlAs or GaAs,
as well as for superlattice, $E_{1}$ and $E_{2}$ are also assigned to be zero
in their studies. The reason why this is incorrect and how they made this
mistake will be discussed in Appendix.

\subsection{Calculations on Nanocrystals Eigenvalue Problems}

\bigskip An atom in a nanocrystal is called the bulk atom when there are 4
first-nearest neighbours and 12 second-nearest neighbours surrounding it,
while an atom that does not fulfill this criterion is called the surface atom.
Accurate and satisfactory interatomic force constants of every atom in a
nanocrystal may be obtained by fully considering the minimum value of the
forces of constraint with arbitrary displacements of atoms in the
nanocrystal\cite{method of relaxation}, known as the \textquotedblleft
relaxation method.\textquotedblright\ Nevertheless, with an optimistic
simplicity one may place the assumption that the interatomic force constants
of Silicon atoms in a nanocrystal may have close values to those in bulk,
except a slight modification at the surface atoms of the nanocrystal. The
assumption is reasonable because in bulk we only consider that a Si atom is
influenced at most by its first and second nearest neighbours, leaving outer
neighbours invisible. Therefore as long as the size of the nanocrystal under
consideration is not too small, the force constants from bulk may still apply
to the nanocrystal.

At the absence of periodicity, the dynamic matrix of a nanocrystal consisting
of $N$ atoms is expanded to a $3N\times3N$ square matrix. The dynamic matrix
is constructed in such a way that each $3$ by $3$ block representing the
mutual force constants between atom $i$ and atom $j$ is placed in positions
designating the rows $(i-1,i,i+1)$ and columns $(j-1,j,j+1)$ in the dynamic
matrix. The $3\times3$ diagonal block placed at the rows $(i-1,i,i+1)$ and
columns $(i-1,i,i+1)$ in the dynamic matrix is understood as the self energy
of the atom $i$, which is discussed below. \bigskip For bulk, the self-energy
of an atom is determined by noticing that the net force exerted on an atom at
the position $(\overset{\rightharpoonup}{R}+\overset{\rightharpoonup}%
{S}_{\sigma})$\thinspace,%
\end{subequations}
\begin{equation}
\sum_{\overset{\rightharpoonup}{R^{^{\prime}}},\sigma^{\prime}}D^{\sigma
\sigma^{^{\prime}}}(\overset{\rightharpoonup}{R}+\overset{\rightharpoonup}%
{S}_{\sigma}-\overset{\rightharpoonup}{R^{^{\prime}}}-\overset{\rightharpoonup
}{S}_{\sigma^{\prime}})\cdot u^{\sigma^{^{\prime}}}(\overset{\rightharpoonup
}{R^{^{\prime}}}) \label{net force on an atom}%
\end{equation}
, vanishes when the atom is at the equilibrium position $\overset
{\rightharpoonup}{d}$. Thus,%
\begin{equation}
\sum_{\overset{\rightharpoonup}{R},\sigma^{\prime}}D^{\sigma\sigma^{^{\prime}%
}}(\overset{\rightharpoonup}{R}+\overset{\rightharpoonup}{S}_{\sigma}%
-\overset{\rightharpoonup}{S}_{\sigma^{\prime}})=0
\label{net force vanishes at d}%
\end{equation}
. This gives%
\begin{equation}
D^{\sigma\sigma}(\overset{\rightharpoonup}{0})=-\sum_{\overset{\rightharpoonup
}{R}\neq\overset{\rightharpoonup}{0},\sigma^{\prime}}D^{\sigma\sigma
^{^{\prime}}}(\overset{\rightharpoonup}{R}+\overset{\rightharpoonup}%
{S}_{\sigma}-\overset{\rightharpoonup}{S}_{\sigma^{\prime}})
\label{self energy}%
\end{equation}
. In other words, the self energy of an atom is the negative sign of the sum
of the 3 by 3 force-constant matrices of the atoms surrounding it. In bulk,
the self energy is thus%
\begin{equation}
D^{\sigma\sigma}(\overset{\rightharpoonup}{0})=-4(A_{\sigma}+2C_{\sigma
}+F_{\sigma})I \label{self energy BULK}%
\end{equation}

Eq \ref{self energy} also holds for nanocrystals by modification of the
equilibrium position for atoms after relaxation. Therefore, for the self
energy of surface atoms that have no 16 ambient atoms in the nanocrystal we
simply sum up the 3 by 3 force-constant matrices of the ambient atoms. It may
be recognized as the free boundary condition after we make the self energy of
surface atoms in this way. The central force assumption up to the
second-neighbor atoms, $E=0$\cite{RIM}, is assumed for surface atoms in order
to preserve symmetry in the dynamic matrix. Afterwards, the standard procedure
of solving the eigenfrequencies and eigenvectors of the dynamic matrix is
applied. Knowledge of group theory regarding to $T_{d}$ group is implemented
to the dynamic matrix. This dramatically ameliorates the requirements of
computer memory and reduces time needed to solve the eigenvalue problem of the
dynamic matrix. Very few number of negative eigenvalues are found because we
did not relax the nanocrystals. For example, 2 out of 3317 modes in the
$T_{2}$ representation of the 7nm case are found to be negative with the
imaginary parts $0.494$ THz and $0.008$ THz. We neglect these kinds of
modes\cite{J Phys Condens Matter 20 145213 2008}.

\subsection{Bond Polarizability Approximation and Intensity of the Raman
Spectrum}

The issue that lies at the heart of the vibrational Raman effects is the
change of polarizability of the nanocrystal on which we measure. The total
polarizability tensor $\alpha_{\rho\sigma}^{tot}$, where $\rho$ and $\sigma$
refer to the Cartesian coordinate index, is readily understood by summing over
every polarizability tensor $\alpha_{\rho\sigma}$($\overset{\rightharpoonup
}{\rho_{ij}}$) connecting atom $i$ and atom $j$:%
\begin{equation}
\alpha_{\rho\sigma}^{tot}=\underset{i<j}{\sum}\alpha_{\rho\sigma}%
(\overset{\rightharpoonup}{\rho_{ij}}) \label{sum polten}%
\end{equation}

Bond polarizability approximation (BPA) model\cite{BPA Bell} suggests that one
may write the polarizability tensor for each bond $\alpha_{\rho\sigma}%
$($\overset{\rightharpoonup}{\rho_{ij}}$) as%
\begin{equation}
\alpha_{\rho\sigma}(\overset{\rightharpoonup}{\rho_{ij}})=\alpha
(\overset{\rightharpoonup}{\rho_{ij}})I+\gamma(\overset{\rightharpoonup}%
{\rho_{ij}})(\overset{\wedge}{\rho_{ij}}\overset{\wedge}{\rho_{ij}}-\frac
{1}{3}I) \label{poltenrij}%
\end{equation}
, where $\alpha(\overset{\rightharpoonup}{\rho_{ij}})$ and $\gamma
(\overset{\rightharpoonup}{\rho_{ij}})$ are isotropic and anisotropic
polarizabilities (called the polarizability parameters) as a function of
position vector associating with atom $i$ and atom $j$, $\overset
{\rightharpoonup}{\rho_{ij}}$, respectively. $I$ is $3\times3$ unit matrix and
$\overset{\wedge}{\rho_{ij}}$ refers to the unit vector of $\overset
{\rightharpoonup}{\rho_{ij}}$. As the nanocrystal is excited by incident light
the characteristic motion of each atom vibrates with a certain eigenfrequency
$w_{l}$ away from its own original equilibrium position, therefore the
position vector between atom $i$ and atom $j$ changes to $\overset
{\rightharpoonup}{\rho_{ij}}=\overset{\rightharpoonup}{r_{ij}}+\overset
{\rightharpoonup}{x_{ij}}$with a slight relative movement $\overset
{\rightharpoonup}{x_{ij}}$ ( more specifically, the relative eigenvectors for
atom $i$ and atom $j$ corresponding to the eigenfrequency $w_{l}$) displacing
away from the relative equilibrium position $\overset{\rightharpoonup}{r_{ij}%
}$. For harmonic vibrations we may expand Eq \ref{poltenrij} with respect to
$\overset{\rightharpoonup}{x_{ij}}$ up to the first order under the assumption
that $\overset{\rightharpoonup}{x_{ij}}\ll\overset{\rightharpoonup}{r_{ij}}$.
First notice that, up to the first order,%
\begin{align}
\overset{\wedge}{\rho_{ij}}\overset{\wedge}{\rho_{ij}}  &  =\frac
{(\overset{\rightharpoonup}{r_{ij}}+\overset{\rightharpoonup}{x_{ij}%
})(\overset{\rightharpoonup}{r_{ij}}+\overset{\rightharpoonup}{x_{ij}}%
)}{\left\vert \overset{\rightharpoonup}{r_{ij}}+\overset{\rightharpoonup
}{x_{ij}}\right\vert ^{2}}\nonumber\\
&  \approx r_{ij}^{-2}\left[  1-\frac{2(\overset{\wedge}{r_{ij}}\cdot
\overset{\rightharpoonup}{x_{ij}})}{r_{ij}}\right]  (\overset{\rightharpoonup
}{r_{ij}}+\overset{\rightharpoonup}{x_{ij}})(\overset{\rightharpoonup}{r_{ij}%
}+\overset{\rightharpoonup}{x_{ij}})
\end{align}
. Secondly we have the expansion for $\alpha(\overset{\rightharpoonup}%
{\rho_{ij}})$ up to the first order as%
\begin{equation}
\alpha(\overset{\rightharpoonup}{r_{ij}}+\overset{\rightharpoonup}{x_{ij}%
})=\alpha(\overset{\rightharpoonup}{r_{ij}})+\alpha^{^{\prime}}(\overset
{\rightharpoonup}{r_{ij}})(\overset{\rightharpoonup}{x_{ij}}\cdot
\overset{\wedge}{r_{ij}})\text{, where }\alpha^{^{\prime}}(\overset
{\rightharpoonup}{r_{ij}})=\frac{\partial\alpha(\overset{\rightharpoonup}%
{\rho_{ij}})}{\partial\overset{\rightharpoonup}{x_{ij}}}|_{\overset
{\rightharpoonup}{\rho_{ij}}=\overset{\rightharpoonup}{r_{ij}}}%
\end{equation}
. Similar formula applies to $\gamma(\overset{\rightharpoonup}{\rho_{ij}})$.
This leads to the first-order expansion of Eq \ref{poltenrij} as the
following:%
\begin{align*}
\alpha_{\rho\sigma}(\overset{\rightharpoonup}{r_{ij}}+\overset{\rightharpoonup
}{x_{ij}})  &  \approx\alpha(\overset{\rightharpoonup}{r_{ij}})+\alpha
^{^{\prime}}(\overset{\rightharpoonup}{r_{ij}})(\overset{\rightharpoonup
}{x_{ij}}\cdot\overset{\wedge}{r_{ij}})+\left[  \gamma(\overset
{\rightharpoonup}{r_{ij}})+\gamma^{^{\prime}}(\overset{\rightharpoonup}%
{r_{ij}})(\overset{\rightharpoonup}{x_{ij}}\cdot\overset{\wedge}{r_{ij}%
})\right]  \cdot\\
&  \left[  \overset{\wedge}{r_{ij}}\overset{\wedge}{r_{ij}}+\frac
{\overset{\wedge}{r_{ij}}}{r_{ij}}\overset{\rightharpoonup}{x_{ij}}%
+\overset{\rightharpoonup}{x_{ij}}\frac{\overset{\wedge}{r_{ij}}}{r_{ij}%
}-\frac{2}{r_{ij}}(\overset{\rightharpoonup}{x_{ij}}\cdot\overset{\wedge
}{r_{ij}})(\overset{\wedge}{r_{ij}}\overset{\wedge}{r_{ij}}+\frac
{\overset{\wedge}{r_{ij}}}{r_{ij}}\overset{\rightharpoonup}{x_{ij}}%
+\overset{\rightharpoonup}{x_{ij}}\frac{\overset{\wedge}{r_{ij}}}{r_{ij}%
})-\frac{I}{3}\right] \\
&  \approx\alpha(\overset{\rightharpoonup}{r_{ij}})+\alpha^{^{\prime}%
}(\overset{\rightharpoonup}{r_{ij}})(\overset{\rightharpoonup}{x_{ij}}%
\cdot\overset{\wedge}{r_{ij}})+\gamma(\overset{\rightharpoonup}{r_{ij}%
})(\overset{\wedge}{r_{ij}}\overset{\wedge}{r_{ij}}-\frac{I}{3})+\gamma
^{^{\prime}}(\overset{\rightharpoonup}{r_{ij}})(\overset{\rightharpoonup
}{x_{ij}}\cdot\overset{\wedge}{r_{ij}})(\overset{\wedge}{r_{ij}}%
\overset{\wedge}{r_{ij}}-\frac{I}{3})\\
&  +\gamma(\overset{\rightharpoonup}{r_{ij}})\left[  \frac{\overset{\wedge
}{r_{ij}}}{r_{ij}}\overset{\rightharpoonup}{x_{ij}}+\overset{\rightharpoonup
}{x_{ij}}\frac{\overset{\wedge}{r_{ij}}}{r_{ij}}-\frac{2}{r_{ij}}%
(\overset{\rightharpoonup}{x_{ij}}\cdot\overset{\wedge}{r_{ij}})\overset
{\wedge}{r_{ij}}\overset{\wedge}{r_{ij}}\right] \\
&  \approx\alpha_{\rho\sigma}(\overset{\rightharpoonup}{r_{ij}})+(\overset
{\rightharpoonup}{x_{ij}}\cdot\overset{\wedge}{r_{ij}})\left[  \alpha
^{^{\prime}}(\overset{\rightharpoonup}{r_{ij}})I+\gamma^{^{\prime}}%
(\overset{\rightharpoonup}{r_{ij}})(\overset{\wedge}{r_{ij}}\overset{\wedge
}{r_{ij}}-\frac{I}{3})\right] \\
&  +\gamma(\overset{\rightharpoonup}{r_{ij}})r_{ij}^{-1}\left[  \overset
{\wedge}{r_{ij}}\overset{\rightharpoonup}{x_{ij}}+\overset{\rightharpoonup
}{x_{ij}}\overset{\wedge}{r_{ij}}-2(\overset{\rightharpoonup}{x_{ij}}%
\cdot\overset{\wedge}{r_{ij}})\overset{\wedge}{r_{ij}}\overset{\wedge}{r_{ij}%
}\right]
\end{align*}
The first term $\alpha_{\rho\sigma}(\overset{\rightharpoonup}{r_{ij}})$ is
neglected because it makes no contribution to Raman effects. Thus the total
polarizability tensor taken into consideration up to the first-order expansion
will be expressed as%
\begin{equation}
\alpha_{\rho\sigma}^{tot}{}^{(1)}=\underset{i<j}{\sum}(\overset
{\rightharpoonup}{x_{ij}}\cdot\overset{\wedge}{r_{ij}})\left[  \alpha
^{^{\prime}}(\overset{\rightharpoonup}{r_{ij}})I+\gamma^{^{\prime}}%
(\overset{\rightharpoonup}{r_{ij}})(\overset{\wedge}{r_{ij}}\overset{\wedge
}{r_{ij}}-\frac{I}{3})\right]  +\gamma(\overset{\rightharpoonup}{r_{ij}%
})r_{ij}^{-1}\left[  \overset{\wedge}{r_{ij}}\overset{\rightharpoonup}{x_{ij}%
}+\overset{\rightharpoonup}{x_{ij}}\overset{\wedge}{r_{ij}}-2(\overset
{\rightharpoonup}{x_{ij}}\cdot\overset{\wedge}{r_{ij}})\overset{\wedge}%
{r_{ij}}\overset{\wedge}{r_{ij}}\right]  \label{first order sum polten}%
\end{equation}
. Similarly the expansion of Eq \ref{poltenrij} up to the second order may
also be derived as%
\begin{align}
\alpha_{\rho\sigma}^{(2)}  &  =%
\frac12
(\overset{\rightharpoonup}{x_{ij}}\cdot\overset{\wedge}{r_{ij}})^{2}\left[
\alpha^{^{\prime\prime}}(\overset{\rightharpoonup}{r_{ij}})I+\gamma
^{^{\prime\prime}}(\overset{\rightharpoonup}{r_{ij}})(\overset{\wedge}{r_{ij}%
}\overset{\wedge}{r_{ij}}-\frac{I}{3})\right] \nonumber\\
&  +\gamma^{^{\prime}}(\overset{\rightharpoonup}{r_{ij}})r_{ij}^{-1}%
(\overset{\rightharpoonup}{x_{ij}}\cdot\overset{\wedge}{r_{ij}})\left[
\overset{\wedge}{r_{ij}}\overset{\rightharpoonup}{x_{ij}}+\overset
{\rightharpoonup}{x_{ij}}\overset{\wedge}{r_{ij}}-2(\overset{\rightharpoonup
}{x_{ij}}\cdot\overset{\wedge}{r_{ij}})\overset{\wedge}{r_{ij}}\overset
{\wedge}{r_{ij}}\right] \nonumber\\
&  +\gamma(\overset{\rightharpoonup}{r_{ij}})r_{ij}^{-2}\left\{
\overset{\rightharpoonup}{x_{ij}}\overset{\rightharpoonup}{x_{ij}}+\left[
4(\overset{\rightharpoonup}{x_{ij}}\cdot\overset{\wedge}{r_{ij}})^{2}%
-x_{ij}^{2}\right]  \overset{\wedge}{r_{ij}}\overset{\wedge}{r_{ij}%
}-2(\overset{\rightharpoonup}{x_{ij}}\cdot\overset{\wedge}{r_{ij}}%
)(\overset{\wedge}{r_{ij}}\overset{\rightharpoonup}{x_{ij}}+\overset
{\rightharpoonup}{x_{ij}}\overset{\wedge}{r_{ij}})\right\}
\label{second order BPA}%
\end{align}
, which involves the second derivative of $\alpha$ and $\gamma$ with respect
to $\overset{\rightharpoonup}{x_{ij}}$. In this study, Eq
\ref{first order sum polten} will be used to calculate the first-order Raman
Spectra for Silicon nanocrystals.

\bigskip For backscattering detection configuration, at which the propagation
direction of the scattered light is parallel but opposite against that of the
(linearly polarized) incident light , \bigskip one may obtain the
temperature-independent (reduced) intensity of the scattered light regarding
to two different polarizations: one for the polarization of the scattered
light parallel to that of the incident light ($I_{s,||}$), the other for the
polarization of the scattered light perpendicular to that of the incident
light ($I_{s,\bot}$). Taken into consideration of averaging over all the
possible orientations of nanocrystals for a given radius, we have\cite{BPA
Bell}$^{\text{,}}$\cite{book D A LONG}:%
\begin{equation}
I_{s,||}\varpropto\underset{l}{\sum}g(w_{l})\cdot(4G_{l}^{2}+45A_{l}^{2})
\label{para intensity}%
\end{equation}
, and%
\begin{equation}
I_{s,\bot}\varpropto\underset{l}{\sum}g(w_{l})\cdot3G_{l}^{2}
\label{per intensity}%
\end{equation}
, where $g(w_{l})$ is the vibrational spectrum and%
\begin{equation}
A_{l}=\frac{1}{3}[\alpha_{11}^{tot}{}^{(1)}+\alpha_{22}^{tot}{}^{(1)}%
+\alpha_{33}^{tot}{}^{(1)}] \label{Al2}%
\end{equation}%
\begin{align}
G_{l}^{2}  &  =%
\frac12
\left\{  [\alpha_{11}^{tot}{}^{(1)}-\alpha_{22}^{tot}{}^{(1)}]^{2}%
+[\alpha_{22}^{tot}{}^{(1)}-\alpha_{33}^{tot}{}^{(1)}]^{2}+[\alpha_{33}%
^{tot}{}^{(1)}-\alpha_{11}^{tot}{}^{(1)}]^{2}\right\} \nonumber\\
&  +3\left\{  [\alpha_{12}^{tot}{}^{(1)}]^{2}+[\alpha_{23}^{tot}{}^{(1)}%
]^{2}+[\alpha_{31}^{tot}{}^{(1)}]^{2}\right\}  \label{Gl2}%
\end{align}

\subsection{Samples Preparations}

\bigskip The silicon nanocrystals are produced under pressures at 0.5 torr, 1
torr, 2 torr, and 4 torr, in which we thermal evaporate the silicon sputtering
target put on a glass substrate that is mounted on the Tantalum (Ta) boat. The
system first is pumped down to 10$^{\text{-3}}$ torr, then in tandem, purged
with Argon gas three times to remove residual water vapor and oxygen. Details
of the fabrication process are described elsewhere\cite{SY lin paper
1}$^{\text{-}}$\cite{SY lin paper 3}. In this situation the silicon atoms form
nanocrystals with various sizes, within some standard deviation, in diameters
of 7 nm, 13 nm, 31 nm, and 37 nm, respectively. The relationship between the
fabrication pressure and the size of Si nanocrystals is discussed by
elsewhere\cite{TED and TEM graphics}. The standard deviations for 7nm and 13nm
cases are both roughly 1nm. There is a clear tendency that under higher
pressure leads to a larger nanocrystal size. This may be explained by noticing
that under higher pressure the higher collision rate is achieved, thus
enhancing the migration and nucleation probability of the Si nanocrystals in
the inert-gas atmosphere. The crystalline property is investigated by
transmission electron diffraction pattern, and nanocrystal sizes are measured
by transmission electron microscopy\cite{TED and TEM graphics}.

\section{Results and Discussions}

\subsection{\bigskip Experimental Details}

Raman Spectra of Silicon nanocrystals with two different radii (7 nm and 13 nm
in diameter) are measured by 488 nm Argon ion laser on Horiba Jobin-Yvon HR800
UV micro-Raman Spectrometer with the groove density of 1200 grooves/mm. The
charge-coupled device detector is cooled down to $-133%
{{}^\circ}%
C$ by liquid nitrogen. A 100\textit{X} objective is used to collect the
back-scattering signals. Exposure time of 100 seconds and accumulation number
of 5 times are required to obtain a clear spectrum in the parallel-polarized
configuration for every sample. The laser power is varied from 5mW to 60 mW
from the source, resulting in the laser power on the surface of samples
ranging from 0.58mW to 5.23mW, which is measured by Thorlabs PM100D optical
power and energy meter.

In order to demonstrate the difference in the spectra for polarization
dependence, we also study the parallel and perpendicular polarizations by
632.8 nm He-Ne laser. Exposure time and accumulation number are set to 30
seconds and 20 times and the laser power on the sample surface is 1.7 mW in
the case. With a resolution of 3.5 cm$^{\text{-1}}$, intensity of the
polarization-dependent scattered light is measured and compared with the
calculation from Eq \ref{para intensity} and Eq \ref{per intensity}. A
standard Silicon chip, treated as the result of bulk Silicon, is also measured
and used to calibrate the system to 520 cm$^{\text{-1}}$.

\subsection{\bigskip Calculations of the Spectra}

\begin{figure}
[t]
\begin{center}
\includegraphics[
trim=4.150838in 0.000000in 0.915693in 0.143375in,
height=6.1514in,
width=4.5429in
]%
{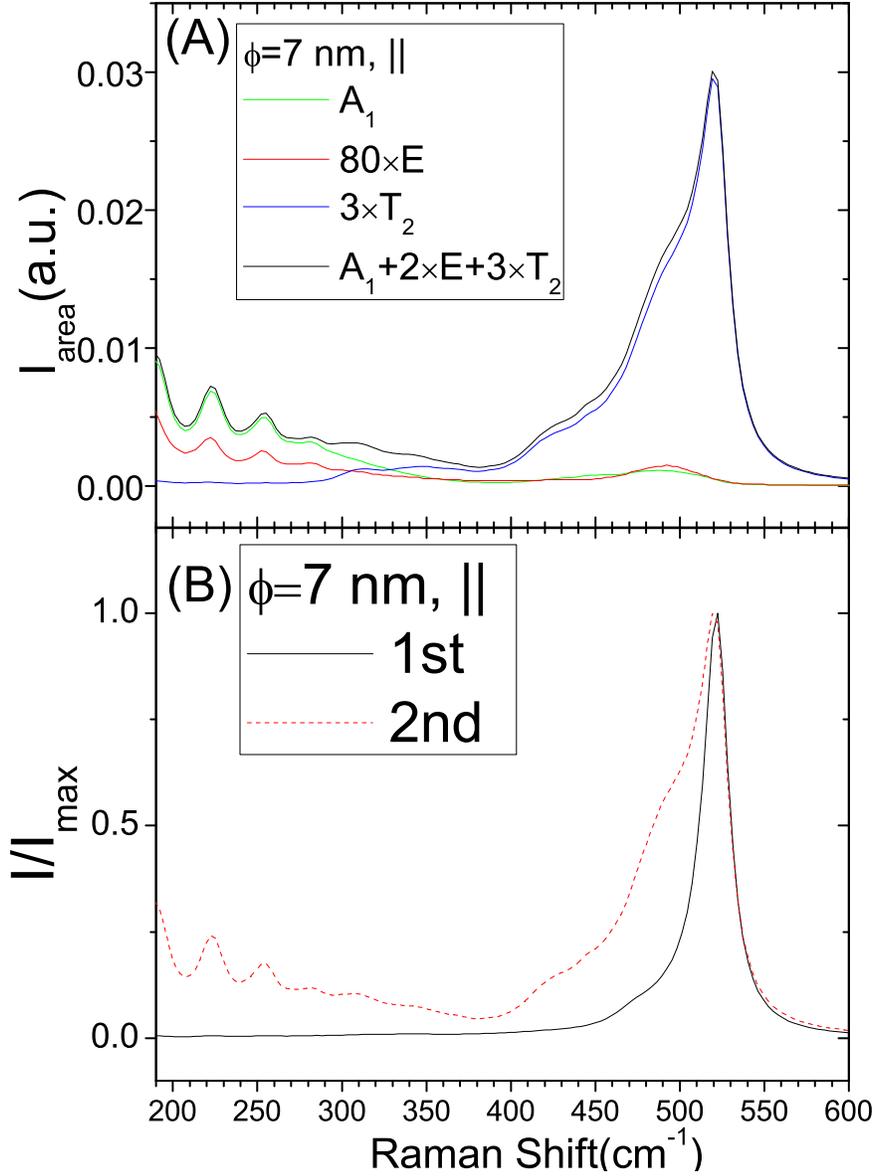}%
\caption{(A) The overall spectrum is the sum of the intensities from the three
Raman-active representations after considering the degeneracy of every
representation. The curve is for the case with the parameters in the bond
polarizability up to the 2nd neighbor. (B) Calculations for parallel
polarization of the first-order Raman spectrum when considering the
first-order and the second-order neighboring atoms in the bond polarizability
parameters. }%
\label{cal diff nb 7 nm}%
\end{center}
\end{figure}

Before comparing the calculations and experiments, there are some remarks of
calculations. First, \bigskip we observe that only intensities involving the
eigenmodes belonging to $T_{2},$ $E,$ and $A_{1}$ are Raman active among the
five irreducible representations $A_{1},$ $A_{2},$ $E,$ $T_{1},$ and $T_{2}%
$\cite{group theory Inui}. Second, the overall spectrum for a nanocrystal with
a specific diameter is acquired by summing over intensities from every
representation and also by taking into consideration the degeneracy of each
representation:%
\begin{equation}
I_{tot}=I_{A_{1}}\times1+I_{E}\times2+I_{T_{2}}\times3
\label{intensity of total spectrum}%
\end{equation}
. Fig \ref{cal diff nb 7 nm}(A) shows how the three Raman-active
representations sum up to a total intensity for the case of parallel
polarization for the 7nm case. In spite of this, to be consistent with the
second-neighbor consideration for the short-range part in the dynamic matrix
when calculating the interatomic force constants, we may also need the
second-neighbor parameters of polarizabilities parameters, $\alpha
(\overset{\rightharpoonup}{r_{ij}})$ and $\gamma(\overset{\rightharpoonup
}{r_{ij}})$. Fig \ref{cal diff nb 7 nm}(B) shows that the spectrum may be
matched well for the asymmetric broadening effect for the parallel
polarization when one considers the polarizability parameters, $\alpha
(\overset{\rightharpoonup}{r_{ij}})$ and $\gamma(\overset{\rightharpoonup
}{r_{ij}})$, up to the second-neighbor atoms. The parameters of the
second-neighbor atoms are chosen to satisfy the ordinary differential
equations of $\alpha(\overset{\rightharpoonup}{r_{ij}})$ and $\gamma
(\overset{\rightharpoonup}{r_{ij}})$\cite{BPA Bell}. This condition, as well
as $\alpha_{c}=\gamma_{c}=0.0001$, is used to calculate the first-order Raman
spectra of nanocrystals composed of 8,597 atoms (7 nm in diameter) and 58,125
atoms (13 nm in diameter). Some anomalous peaks in the range 200
cm$^{\text{-1}}$ to 300 cm$^{\text{-1}}$ are shown in Fig
\ref{cal diff nb 7 nm} ( 221 cm$^{\text{-1}}$, 254 cm$^{\text{-1}}$, and 281
cm$^{\text{-1}}$ ) but are not shown in the experiments because they are
mainly due to single-size effects thus being smeared out by size deviation in
real samples. At last, taking into account the nature of shape of the
nanocrystals and the fact that the surface of every nanocrystal is
hydrogenated, both resulting in the smaller values of eigenfunctions that
stand for the motion of the atoms on the surface, we find that the experiments
are well described when we divide the eigenfunctions in the $A_{1}$ and $E$
representations, whose eigenvalues are smaller than 300 cm$^{-1},$ by 8 for
the 7nm case while dividing by 6 for the 13 nm case.

\subsection{Comparisons}

\subsubsection{\bigskip The Spectra measured by He-Ne Laser}

\begin{figure}
[t]
\begin{center}
\includegraphics[
trim=4.855568in 0.000000in 0.482364in 0.206320in,
height=6.2716in,
width=4.4849in
]%
{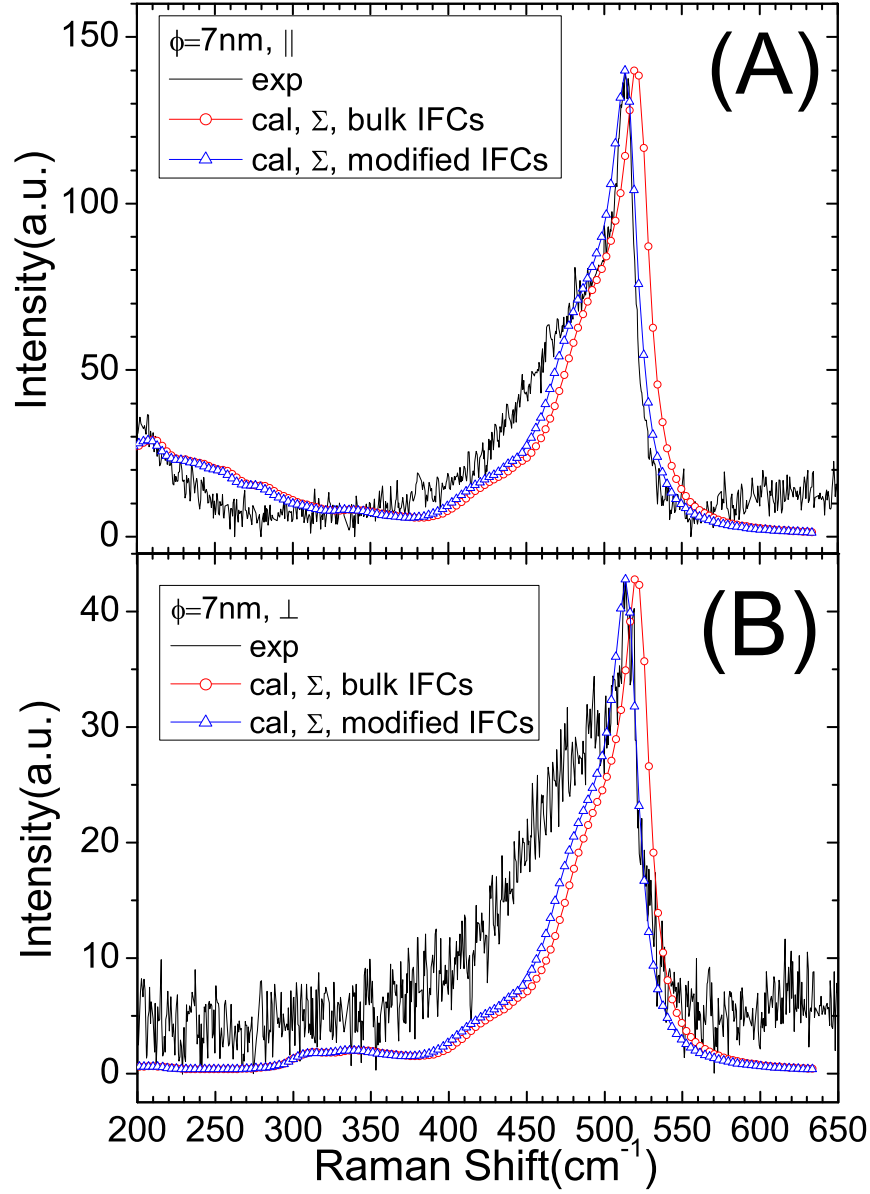}%
\caption{Comparisons of the first-order Raman spectra between
calculations(cal) and He-Ne experiments(exp) under parallel ($\vert$$\vert$,
Fig A) and perpendicular ($\bot$, Fig B) polarizations for nanocrystals of 7
nm in diameter. Single-size peaks from 200 cm$^{\text{-1}}$ to 300
cm$^{\text{-1}}$ are smeared out after summing up all the spectra of the size
deviations ranging from 6.1 nm to 8.4 nm. (cal, $\sum$). After assigning
A=-9.56431, we may get correct peak position. Because Intensity from
perpendicular-polarized A$_{1}$ modes is zero, the curve is flattened out in
the low frequency region in Fig B.}%
\label{cal vs exp 7nm}%
\end{center}
\end{figure}

\begin{figure}
[t]
\begin{center}
\includegraphics[
trim=5.061969in 0.000000in 0.000000in 0.000000in,
height=7.6285in,
width=5.5573in
]%
{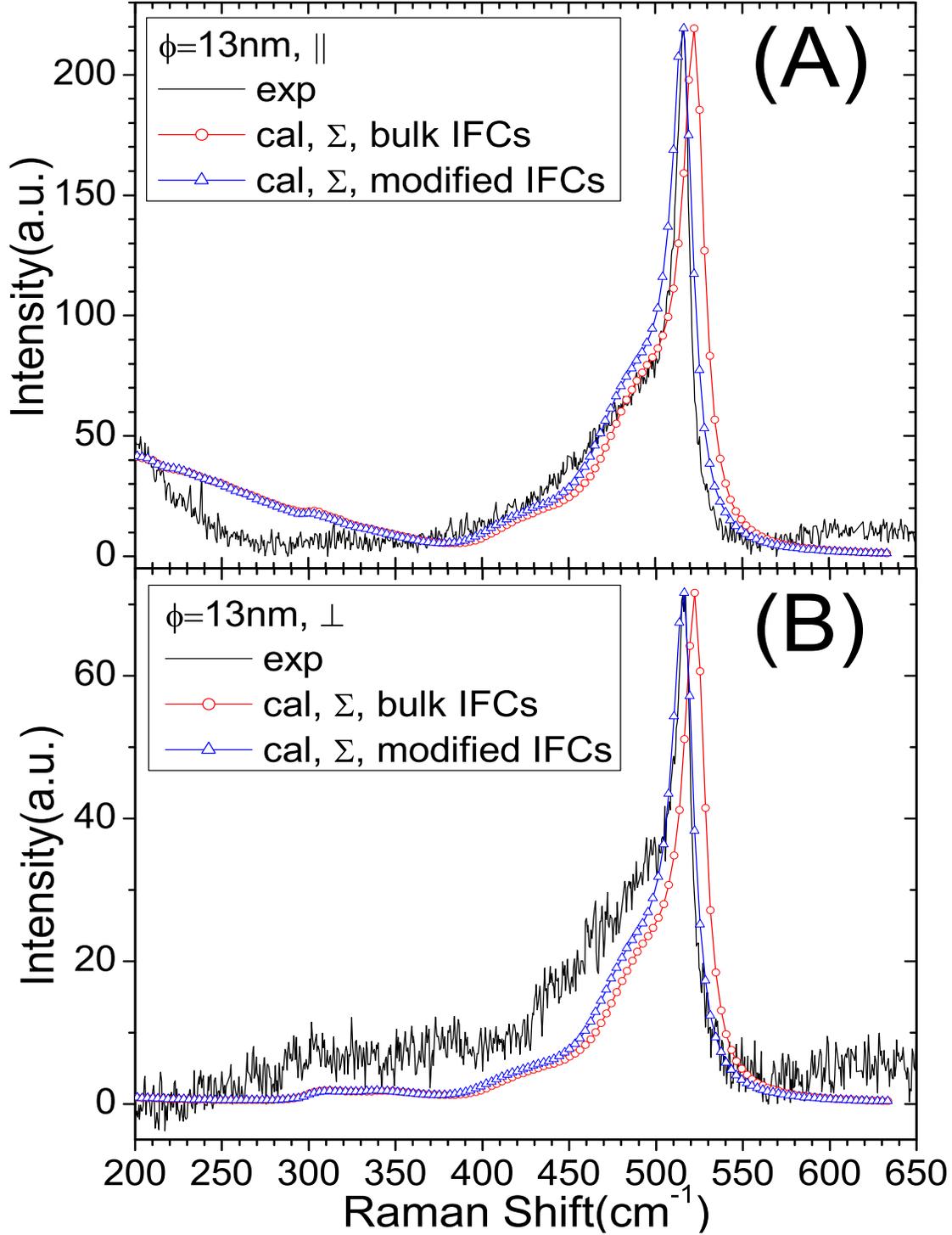}%
\caption{Comparisons of the first-order Raman spectra between
calculations(cal) and experiments(exp) under parallel ($\vert$$\vert$, Fig A)
and perpendicular ($\bot$, Fig B) polarizations for nanocrystals of 13 nm in
diameter. After assigning A=-9.58383, we may get correct peak position.}%
\label{cal vs exp 13nm}%
\end{center}
\end{figure}

Comparisons with experiments measured by He-Ne laser with satisfying
similarities in between are shown in Fig \ref{cal vs exp 7nm} and Fig
\ref{cal vs exp 13nm} for the polarization-dependent study. The three
anomalous peaks shown in Fig \ref{cal diff nb 7 nm} are smeared out after
considering the standard deviation ($\sim$1nm) of the 7nm nanocrystal
diameters. Notice that when using bulk IFCs, the DOS of the nanocrystal may
almost reproduce that of bulk when the nanocrystal size is larger than 6nm,
but one immediately recognizes that the mode at 320 cm$^{\text{-1}}$ shown in
the DOS of nanocrystal does not contribute to the first-order Raman spectrum
and that the one-phonon effect makes no significant contribution to the lines
around 300 cm$^{\text{-1}}$. Therefore it should be totally due to the
two-phonon effect, being recognized as 2TA\cite{PRB 64 073304}$^{\text{,}}%
$\cite{two phonons in si bulk}. For both cases, the DOS curves have the peak
at 488 cm$^{\text{-1}}$ while the calculated Raman spectra show the 1LO peak
at 521 cm$^{\text{-1}}$. Because modes in the $A_{1}$ representation lead to
essentially zero intensity in Eq \ref{per intensity} , we see that in Fig
\ref{cal vs exp 7nm} (B) and Fig \ref{cal vs exp 13nm} (B) the spectra with
perpendicular polarization are flattened out in the 200 cm$^{\text{-1}}$ to
350 cm$^{\text{-1}}$ region. Besides, the peak positions may be obtained in
calculation by assigning appropriate IFCs. For example, in the 7nm case, we
may put $A$=$-9.56431$, while in the 13nm case, we may set $A=-9.58383$. For
the rest of the IFCs we assign the values with the same ratio of change
between bulk value $A$ and the assigned new value $A$.

\begin{figure}
[t]
\begin{center}
\includegraphics[
trim=0.378593in 0.680159in 1.433407in 0.357564in,
height=5.4405in,
width=6.7628in
]%
{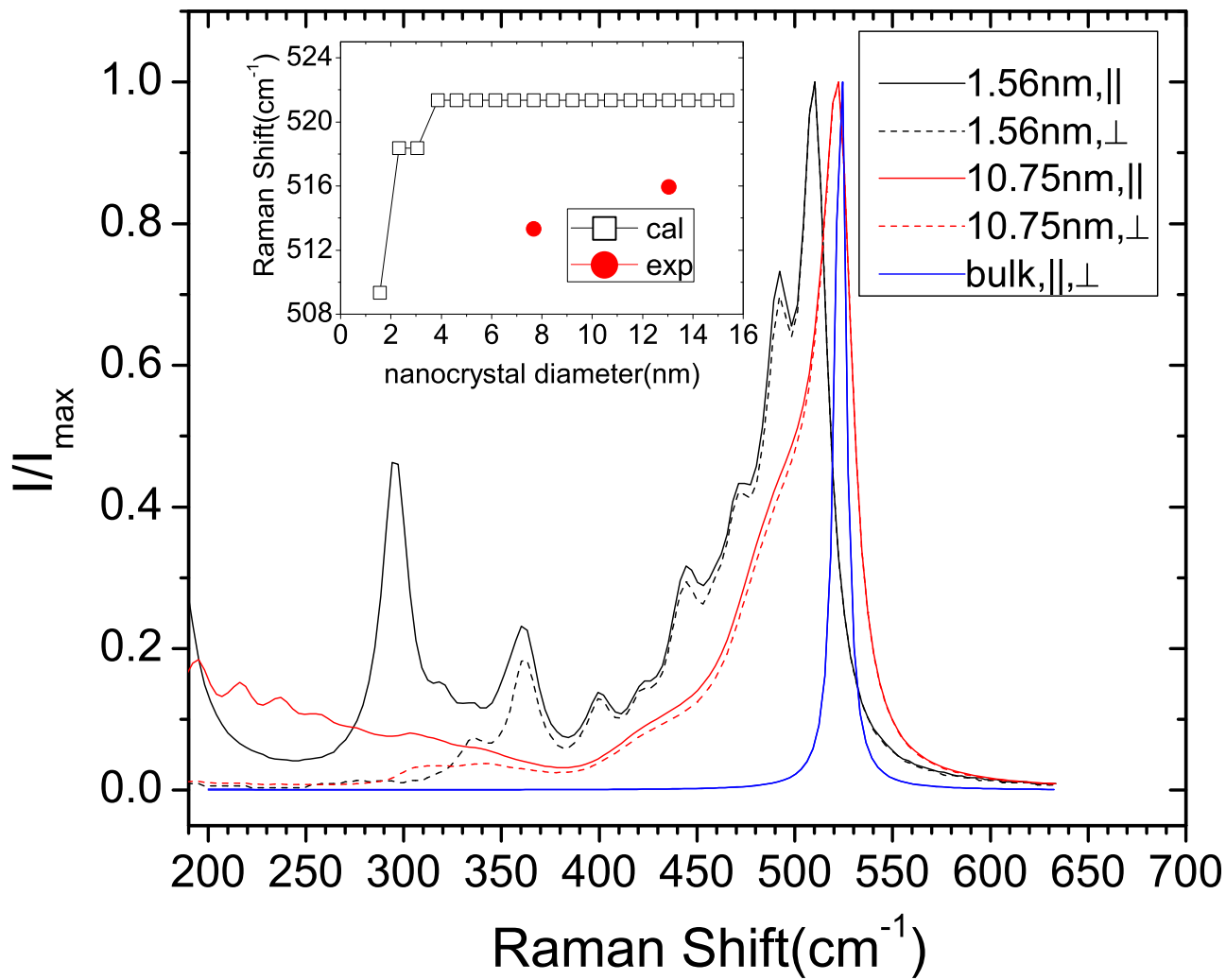}%
\caption{Calculated Raman Shifts vs. nanocrystal diameters with bulk IFCs. The
1.56 nm nanocrystal shows the peak at 509 cm$^{\text{-1}}$ while the 10.75 nm
one at 521 cm$^{\text{-1}}$. Bulk has a shift of 524 cm$^{\text{-1}}$. (Inset)
The calculated results, shown in square, overestimate the experimental ones,
shown in circle.}%
\label{1 LO peak shift vs size}%
\end{center}
\end{figure}

To demonstrate the redshifts for smaller nanocrystals, we use the bulk IFCs
and compare the Raman spectra in Fig \ref{1 LO peak shift vs size} for both
polarization directions for nanocrystals with diameters 1.56 nm and 10.75 nm
along with the bulk which shows the delta-like spectrum with peak at 524
cm$^{\text{-1}}$. In the inset a clear redshift is shown from 524
cm$^{\text{-1}}$ down to 509 cm$^{\text{-1}}$. One observes that the
calculated Raman shifts are overestimated ( in the value of 4-7 cm$^{\text{-1}%
}$) compared with those from experiments\cite{PRB 73 033307 2006}. There are
some aspects in this regard.

The first account for this inconsistency may be that the measured nanocrystals
are not uniform in size and we should consider along with the standard
deviation of the sizes. But study shows that the standard deviation for 7nm
and 13nm cases is roughly 1nm, and in the region the Raman shift barely
changes according to the inset in Fig \ref{1 LO peak shift vs size}. This
indicates that even if we consider the size effects of samples due to standard
deviation from the production process it will not improve the peak shifts to
get closer to the experimental results. Another possible reason may be due to
the local heating caused by the laser spot which is roughly 2 or 3 $\mu m$ in
diameter and has intensity of several milliWatts on the samples\cite{PRB 80
193410 2009}$^{\text{,}}$\cite{PRB 66 161311R 2002 local heating}. Observable
peak shifts are also claimed to be found when tuning the laser to higher power
on the same sample. Therefore some studies point out that we may overemphasize
the effects of quantum confinements on the peak shift.

\subsubsection{The Spectra measured by 488nm Argon-ion Laser}

\begin{figure}
[t]
\begin{center}
\includegraphics[
trim=5.977663in 0.000000in 0.000000in 0.000000in,
height=7.0681in,
width=5.2572in
]%
{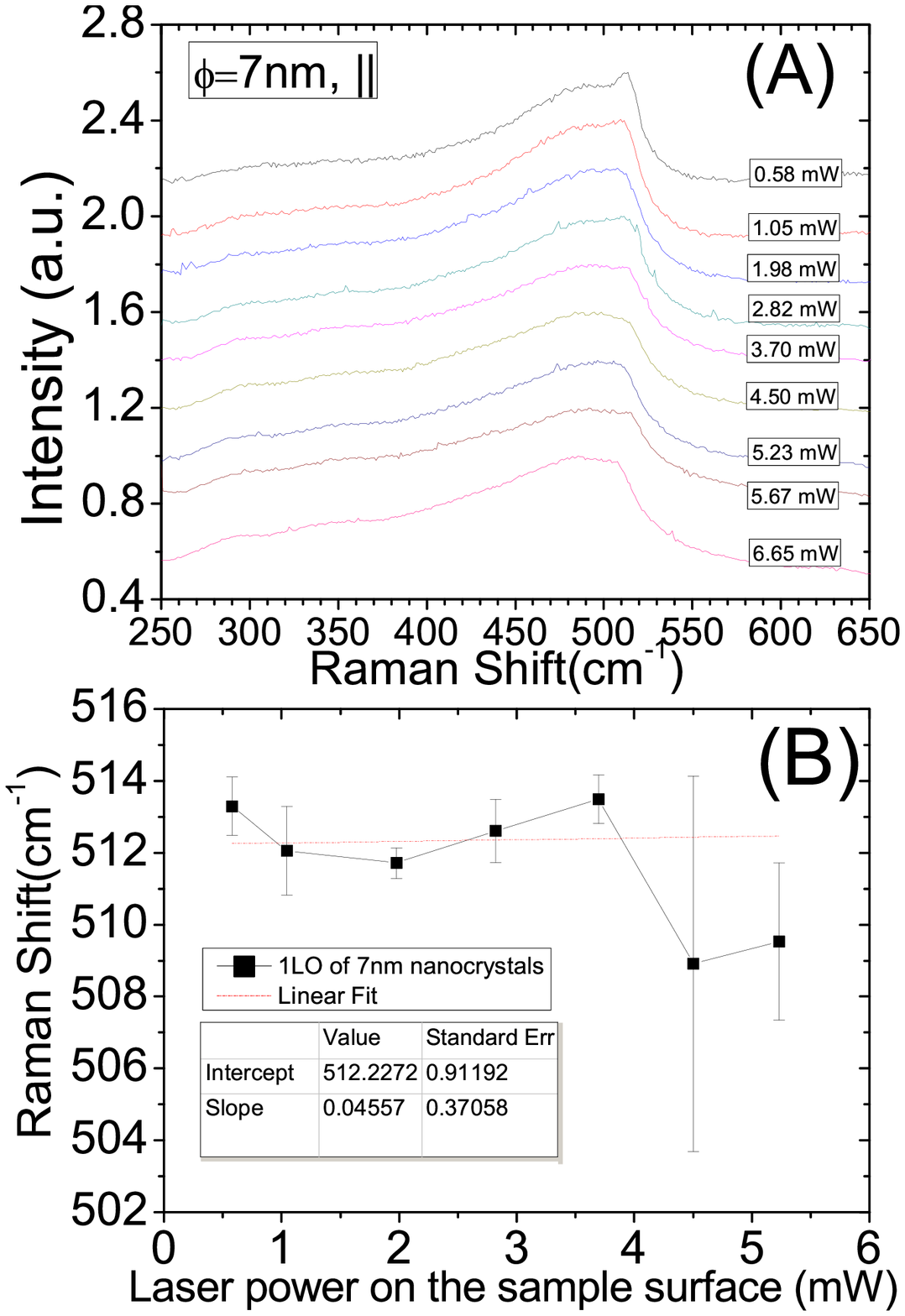}%
\caption{Power variations by 488nm Argon-ion laser on the 7nm case. (A) Raman
Spectra under different laser powers on the sample surface. (B) 1LO Raman
Shift vs. laser power on the sample surface. No significant shift is observed
under different laser powers.}%
\label{GRAPH_local_heating_7nm}%
\end{center}
\end{figure}

\begin{figure}
[t]
\begin{center}
\includegraphics[
trim=2.735676in 0.000000in 2.737957in 0.000000in,
height=7.7366in,
width=5.2546in
]%
{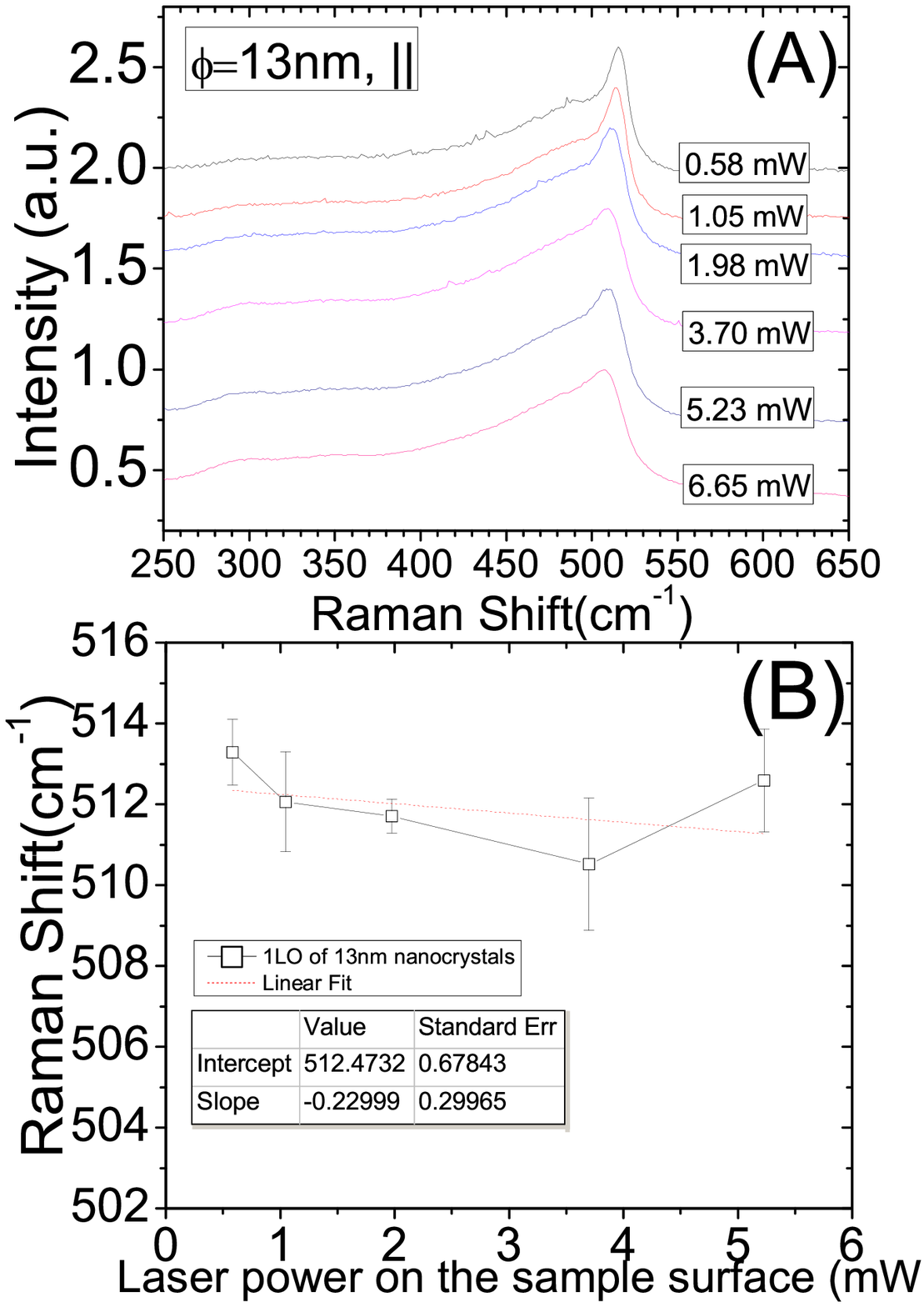}%
\caption{Power variations by 488nm Argon-ion laser on the 13nm case. (A) Raman
Spectra under different laser powers on the sample surface. (B)1LO Raman Shift
vs. laser power on the sample surface. No significant shift is observed under
different laser powers.}%
\label{GRAPH_local_heating_13nm}%
\end{center}
\end{figure}

In order to investigate this, spectra changes of parallel-polarized scattered
light under various incident laser powers in the 7nm and 13nm cases are
measured by 488nm Argon-ion laser. To obtain the standard deviation, under
each laser power the experiment is repeated 5 times. However, we see that, in
Fig \ref{GRAPH_local_heating_7nm}(B) and Fig \ref{GRAPH_local_heating_13nm}%
(B), the 1LO peak shift barely changes under different laser powers. This is
contradictory to the studies\cite{PRB 80 193410 2009}$^{\text{,}}$\cite{PRB 66
161311R 2002 local heating}. The study\cite{PRB 80 193410 2009} points out
that peak shift influenced by the quantum confinement effect should be
negligible for nanocrystals larger than 6 nm, which may be shown by our
calculation in the inset of Fig \ref{1 LO peak shift vs size} when using bulk
IFCs. However, extrapolation in Fig \ref{GRAPH_local_heating_7nm}(B) and Fig
\ref{GRAPH_local_heating_13nm}(B) shows that under zero laser power the Raman
shift should be both around 512 cm$^{\text{-1}}$ for 7nm and 13nm. Therefore
we believe that the quantum confinement still plays the most important role in
the redshift (because relaxing the nanocrystal may lead to correct IFCs) for
the nanocrystal at least with the size up to 13nm.

Raman peak shifts measured by He-Ne laser for 7nm and 13nm are 513
cm$^{\text{-1}}$ and 516 cm$^{\text{-1}}$, respectively. Compared with the
zero-power peak shift, 512 cm$^{\text{-1}}$ for both 7nm and 13nm cases, they
are both within the system measuring error. This indicates that the spectra in
Fig \ref{cal vs exp 7nm} and Fig \ref{cal vs exp 13nm} by 1.7mW He-Ne laser
could be thoroughly ascribed to the quantum confinement effect; the laser
local heating may not contribute in the case. In addition, we may infer that
the laser local heating may not be very important for the sizes as long as the
power is lower than 1.7mW.

\subsection{Surface Effects on Some Specific Modes}

\begin{figure}
[t]
\begin{center}
\includegraphics[
trim=0.586135in 0.380294in 0.936219in 0.928443in,
height=4.862in,
width=6.4524in
]%
{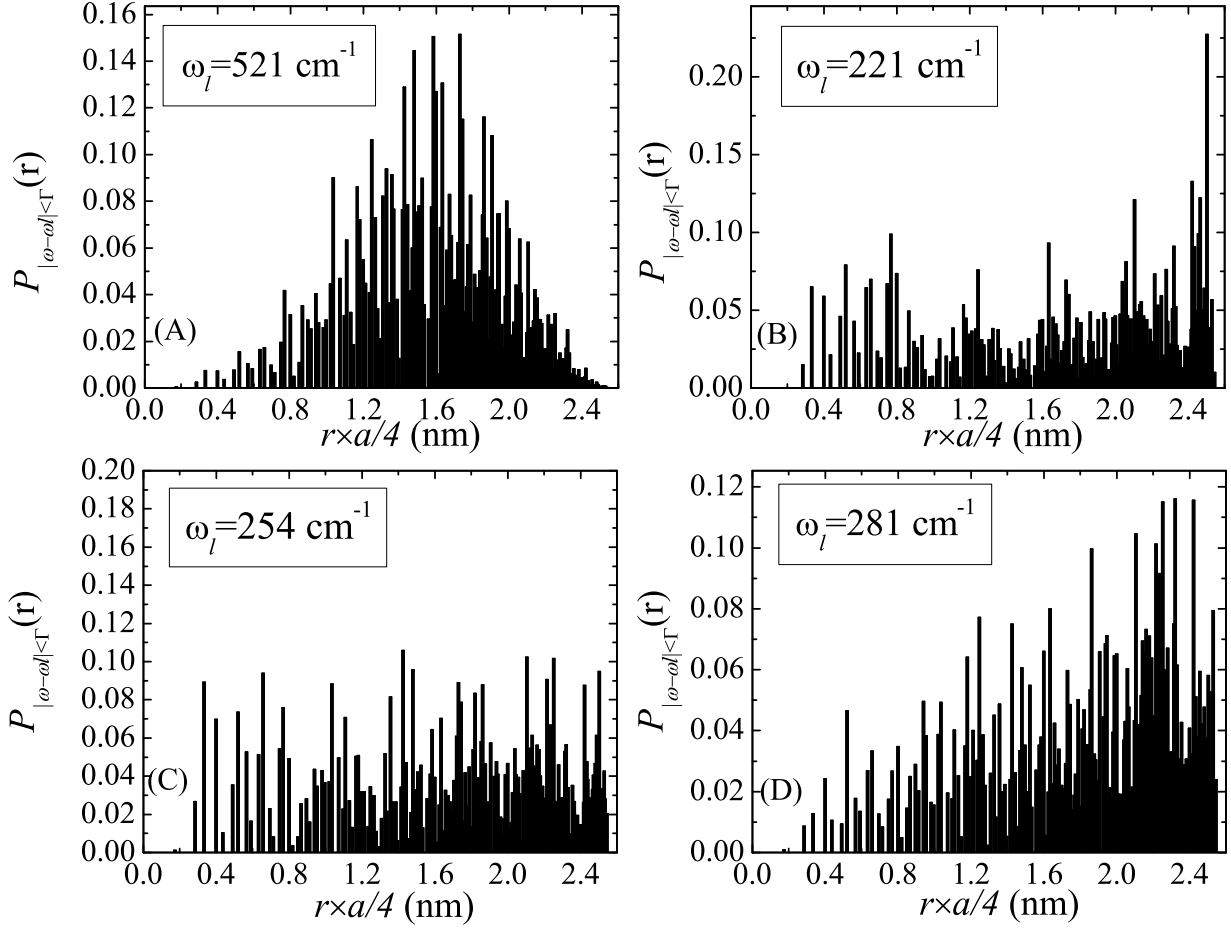}%
\caption{Radial probability distributions for the 7nm case. (A) Bulk-like
Raman shift. From (B), (C), and (D) we calculate that the ratios of integrated
probability are 40\%, 22\%, and 33\%, respectively. $a=5.43$.}%
\label{probability distribution}%
\end{center}
\end{figure}

We now analyze the effects from surface atoms on the three anomalous peaks for
the 7nm case observed in Fig \ref{cal diff nb 7 nm} but not shown in
experiments. For comparisons we also include the effect on the bulk-like peak,
521 cm$^{\text{-1}}$. To begin with, we study the origin of the four peaks by
defining the \textit{radial probability distribution }$P_{|w-w_{l}|<\Gamma
}(r)$ for a certain eigenfrequency $w_{l}$, within a natural broadening
$\Gamma$, at the distance $r$ away from the center of the nanocrystal with the
radius $R$ as%

\begin{equation}
P_{|w-w_{l}|<\Gamma}(r)=\underset{j\in r}{\sum}(\overset{\wedge}{x_{0j}}%
\cdot\overset{\wedge}{x_{0j}}) \label{radial probability distribution}%
\end{equation}
, where $\overset{\wedge}{x_{0j}}$ refers to the normalized eigenvector of the
atom $j$. Only modes belonging to $A_{1}$, $E$, and $T_{2}$ are summed in a
way similar to Eq \ref{intensity of total spectrum}. To see the effects from
surface atoms we may calculate the ratio of integrated probability between
surface atoms and bulk atoms:%

\begin{equation}
\Omega_{w_{l}}=\frac{\int_{r\in S}P_{|w-w_{l}|<\Gamma}(r)dr}{\int_{r\in
B}P_{|w-w_{l}|<\Gamma}(r)dr} \label{integration probability}%
\end{equation}
, where $S$ and $\ B$ refer to surface atoms and bulk atoms, respectively. In
the case of the 7nm nanocrystals, surface atoms refer to atoms in the range of
$r=2.26$ to $r=2.55$. From Fig \ref{probability distribution} we obtain
$\Omega_{w_{l}=521}=3\%$, $\Omega_{w_{l}=221}=40\%$, $\Omega_{w_{l}=254}%
=22\%$, and $\Omega_{w_{l}=281}=33\%$. Therefore, surface atoms barely have an
effect on the bulk-like 521cm$^{\text{-1}}$ peak while they may have an
effect, at most $40\%$, on the other peaks from single-size contributions. Our
samples in reality may be squeezed ellipsoids and this may account for at
least $20\%$ why the three modes arising partly from surface atoms may not be
shown in experiments. Nonetheless, as discussed above, the most important
factor is that the size deviation in real samples smears out the single-size peaks.

\section{Conclusions}

\bigskip In conclusions, we first obtain the interatomic force constants up to
the second-neighbor atoms, under the rigid-ion model, by directly solving
multiple equations that are eigenvalues of the dynamic matrix for bulk Si at
the three special points\ $\Gamma,X,$ and $L$ in the Brillouin zone. The six
interatomic force constants (IFCs) thus obtained are compared with the results
from ab initio calculations. Besides matching well for the two sets of
parameters, we may include as many IFCs for more neighboring atoms as we
desire by implementing the results from ab initio calculations. The
characteristic vibrations of nanocrystals for free boundary conditions are
solved by assuming that the IFCs among atoms in bulk are optimistically
applicable \bigskip to the atoms in the nanocrystals, with a slight
modification of the self energy for surface atoms.

Nanocrystals with sizes of 7nm and 13nm in diameter are produced by thermal
evaporations and measured in the backscattering configuration on Raman
spectrometer with 632.8nm He-Ne laser and 488nm Argon-ion laser. The
polarization of scattered light perpendicular or parallel to that of the
incident light is distinguished by using an analyzer in front of the confocal
hole. Besides, the first-order Raman spectra are calculated by bond
polarizability approximation model. In order to get more satisfactory
comparison in the spectrum compared with experiments, the two polarizability
parameters are extended up to the second-neighbor atoms. Under this
circumstance, asymmetric broadening is seen in fairly good resemblance between
calculations and experiments for both sizes measured by the He-Ne laser.
Moreover, to investigate the redshifts, various sizes of nanocrystals ranging
from 1.56 nm to 15.36 nm are calculated. We see that when using the same IFCs
as bulk in the nanocrystals, the peak shifts are almost the same as the
diameter of nanocrystals is larger than 4nm; the calculated redshifts
overestimate the experimental results at the amount of 4-7 cm$^{\text{-1}}$,
which looks like influenced by the paser local heating effect. But study on
laser-power dependence show no significant shifts and therefore it is dubious
about the effect of local heating on the redshifts in the Raman Spectrum. We
believe that the local heating at most affects the asymmetric broadening on
the spectrum shoulders, for higher intensity generates more vibrational modes.
However, the peak position should not be influenced. Therefore we infer that
the quantum confinement effect plays the major role in the peak shift. Our
study shows that we may acquire close peak shift after assigning proper values
of IFCs whose change is due to the quantum confinement. For example, in the
7nm case the value $A$ in IFCs changes to $A=-9.56431$, while in the 13nm
case, we may set $A=-9.58383$. Extrapolation in the 488nm laser-power
dependence shows that both 7nm and 13nm cases have Raman shift around 512
cm$^{\text{-1}}$ under zero laser power at which the peak shift is totally due
to the quantum confinement effect. Compared with the shifts in the He-Ne
study, 513 cm$^{-1}$ for the 7nm case and 516 cm$^{-1}$ for the 13nm case, and
noting that each of the two values is within the measuring error of the
zero-power shift, 512 cm$^{-1}$, it is clear that in the He-Ne study when the
laser power is 1.7mW, local heating effect contributes nothing to the redshift.

Finally, the ratios of the integrated probability between surface atoms to
bulk atoms show that surface atoms contribute only 3\%\ for the bulk-like mode
while they make more contributions, from 20\%\ to 40\%, to the other modes
originating from nanocrystals in single size.

\begin{acknowledgments}
The author thanks for the fellowship support from Taiwan International
Graduate Program, Academia Sinica, Taiwan, Republic of China. We also thank
for Dr. R. Thangavel and Dr. Hsu Shih-Hsin for useful discussions.
\end{acknowledgments}

\section{\bigskip Appendix}

Some publications\cite{chang wrong paper 1}$^{-}$\cite{chang wrong paper 3}
set $E_{1}$ and $E_{2}$ to be zero, however, they did not give any physical
explanation. As a matter of fact, one may obtain from the ab initio
calculation that neither $E_{1}$ nor $E_{2}$ is zero. Indeed, for some
materials $E_{1}$ and $E_{2}$ are zero due to symmetry property. Rocksalt
structure, such as NaCl, is an example, but not for the zincblende structure.
The problem lies in the wrong arrangement in the short-range part of the
dynamic matrix when looking for the three dimensional force-constant matrices
other than that of the atom at the position $\frac{a}{4}$($2,2,0$) for which
we define in Eq \ref{DM_2nd_nb}. They chose wrong similarity transformation
$T_{i}$, resulting in incorrect symmetry in the short-range part of the
dynamic matrix. Every element in the dynamic matrix involving $E_{1}$ and
$E_{2}$ is wrong. The correct elements involving $E_{1}$ and $E_{2\text{ }}$in
the dynamic matrix\cite{kunc RIM CPC} for calculating zincblende phonon band
structure are listed in the row labeled as \textquotedblleft
correct\textquotedblright\ in Table II. Terms conjugating to those in Table
II\ are also required to be taken into account. However, in the
calculations\cite{chang wrong paper 1}$^{-}$\cite{chang wrong paper 3} their
counterparts are obviously inconsistent. This is due to incorrect assignment
for the 3$\times$3 force-constant matrices of the second-nearest atoms.

\begin{figure}
[t]
\begin{center}
\includegraphics[
trim=4.980662in 1.394463in 3.309459in 0.802728in,
height=6.4671in,
width=3.4731in
]%
{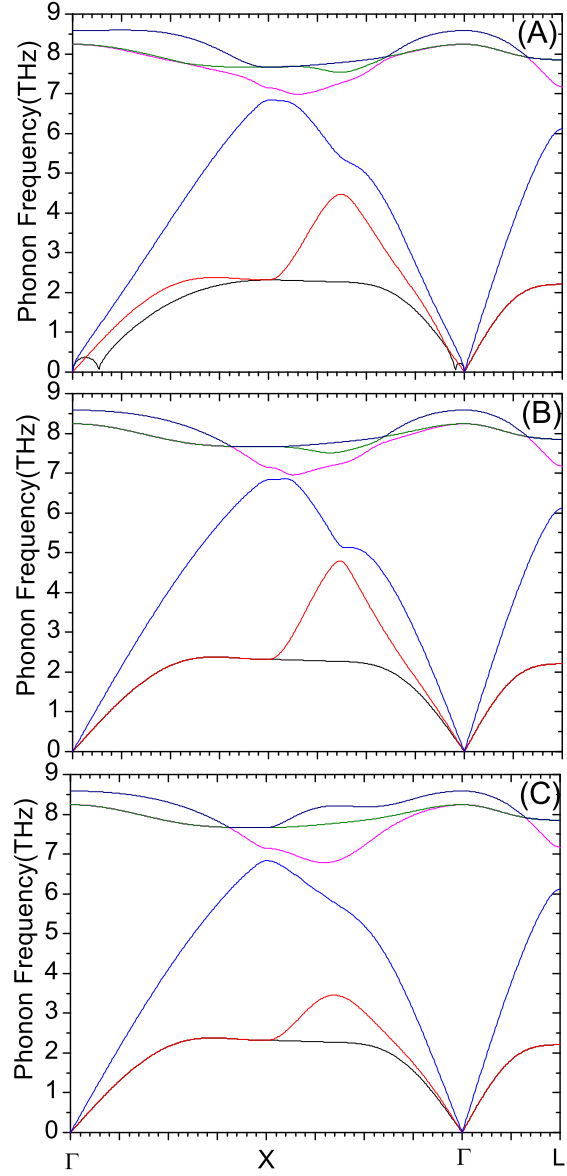}%
\caption{GaAs phonon band structure. (A) When setting nonzero E$_{1}$ and
E$_{2}$ in their original fortran procedure, an unpleasant bump is shown near
the zone center owing to the wrong symmetry in the short-range part in the
dynamic matrix. (B) When assigning the E$_{1}$ and E$_{2}$ to be zero, the
unwanted bump is removed, but we have incorrect optical branches while
comparing with the correct band structure in (C).}%
\label{Appen GaAs}%
\end{center}
\end{figure}

This mistake seems indifferent for structures with $E_{1}$ and $E_{2}$ to be
zero, such as NaCl or MgO. But for materials in the diamond structure, such as
Silicon, the mistake may draw to a misleading conclusion that
\textquotedblleft for Group IV, we must have $E_{1}$ $=$ $E_{2}$ $=0$ to
preserve symmetry.\textquotedblright\ In fact, symmetry property in the
diamond structure does not restrict the two terms to be zero. Kunc, et.
al.\cite{kunc previous}, once stated that $E_{1}$ $=$ $E_{2}$ $=0$ when only
considering the central-force assumption, but it has nothing to do with
symmetry property consideration. Furthermore, there is no reason to take
$E_{1}$ $=$ $E_{2}$ $=0$ for materials in zincblende or superlattice, but
while setting $E_{1}$ and $E_{2}$ to be nonzero, the lethal mistake will give
an unpleasant \textquotedblleft\textit{bump}\textquotedblright\ near the zone
center when exploiting the parameters\cite{J Phys Conden Matt 2 1457 1990} for
GaAs, which is shown in Fig. \ref{Appen GaAs} (A).

To avoid the bump, one may mistakenly replace $E_{1}$ and $E_{2}$ to be zero,
with the phonon band structure shown in Fig. \ref{Appen GaAs}(B). But it does
not give the correct structure for optical branches when comparing with the
correct ones with appropriate arrangement in the short-range part shown in
Fig. \ref{Appen GaAs}(C).

However, it raises a very serious question on whether or not their original
arrangement in the short-range part produces correct superlattice phonon band
structure. Compared with the correct short-range part in the dynamic matrix it
is easy to find out that the optical branches of the superlattice in their
publications are essentially incorrect, because they did not get correct
optical branches for GaAs after assigning E$_{1}$ and E$_{2}$ to be zero.
Therefore, it is imperative to examine again carefully on all of their
superlattice-related publications with the rigid-ion model.%

\bigskip%

%

%

%

%

%

%

%

\begin{table}[ptb]
\caption{IFCs obtained either by solving Eq. 5 and Eq. 6 with neutron
scattering data or by Abinit calculation.}%
\centering%
\begin{tabular}
[c]{ccccccc}\hline\hline
& $A$ & $B$ & $C$ & $F$ & $D$ & $E$\\\cline{3-5}%
Neutron Scattering Data\footnote{In unit of THz.} & 15.69 &
\multicolumn{3}{c}{%
\begin{tabular}
[c]{c}%
13.69\\
12.45\\
4.51
\end{tabular}
} &
\begin{tabular}
[c]{c}%
14.62\\
12.51\\
10.97\\
3.41
\end{tabular}
& \\\hline
Solutions from Eq. 5 and Eq. 6\footnote{In unit of kg/sec$^{2}$ after each
parameter value is multiplied by $5.7641$.} & -9.8315\footnote{At $\Gamma$.} &
-6.6725 & -0.6373 & 1.4045\footnote{At X by solving Eq. 5.} &
-0.5210\footnote{At $L$. $D$ is chosen to get the closest fit to Eq. 6.} &
0.0189\footnote{$E$ is chosen to fit the values in the L--X--W--L direction.
}\\
Abinit calculations$^{\text{b}}$ & -9.1264 & -6.4525 & -0.4970 & 1.1317 &
-0.4781 & 0.2944\\\hline\hline
\end{tabular}
\end{table}

\bigskip

\bigskip

\begin{table}[b]
\caption{Comparisons between correct and wrong elements involving the
parameters E$_{1}$ and E$_{2}$ in the dynamic matrix for the short-range
part.}%
\centering%
\begin{tabular}
[c]{cc}\hline\hline
$DM_{[1,2]}$ &
\begin{tabular}
[c]{ccc}%
correct &  & $-4D_{1}\sin(\pi k_{1})\sin(\pi k_{2})+4iE_{1}\sin(\pi k_{3}%
)\cos(\pi k_{1})-4iE_{1}\sin(\pi k_{3})\cos(\pi k_{2})$\\
wrong &  & $-4D_{1}\sin(\pi k_{1})\sin(\pi k_{2})+4iE_{1}\sin(\pi k_{1}%
)\cos(\pi k_{3})-4iE_{1}\sin(\pi k_{2})\cos(\pi k_{3})$%
\end{tabular}
\\\hline
$DM_{[1,3]}$ &
\begin{tabular}
[c]{ccc}%
correct &  & $-4D_{1}\sin(\pi k_{1})\sin(\pi k_{3})+4iE_{1}\sin(\pi k_{2}%
)\cos(\pi k_{1})-4iE_{1}\sin(\pi k_{2})\cos(\pi k_{3})$\\
wrong &  & $-4D_{1}\sin(\pi k_{1})\sin(\pi k_{3})+4iE_{1}\sin(\pi k_{1}%
)\cos(\pi k_{2})-4iE_{1}\sin(\pi k_{3})\cos(\pi k_{2})$%
\end{tabular}
\\\hline
$DM_{[2,3]}$ &
\begin{tabular}
[c]{ccc}%
correct &  & $-4D_{1}\sin(\pi k_{2})\sin(\pi k_{3})+4iE_{1}\sin(\pi k_{1}%
)\cos(\pi k_{2})-4iE_{1}\sin(\pi k_{1})\cos(\pi k_{3})$\\
wrong &  & $-4D_{1}\sin(\pi k_{2})\sin(\pi k_{3})+4iE_{1}\sin(\pi k_{2}%
)\cos(\pi k_{1})-4iE_{1}\sin(\pi k_{3})\cos(\pi k_{1})$%
\end{tabular}
\\\hline
$DM_{[4,5]}$ &
\begin{tabular}
[c]{ccc}%
correct &  & $-4D_{2}\sin(\pi k_{1})\sin(\pi k_{2})-4iE_{2}\sin(\pi k_{3}%
)\cos(\pi k_{1})+4iE_{2}\sin(\pi k_{3})\cos(\pi k_{2})$\\
wrong &  & $-4D_{2}\sin(\pi k_{1})\sin(\pi k_{2})+4iE_{2}\sin(\pi k_{1}%
)\cos(\pi k_{3})-4iE_{2}\sin(\pi k_{2})\cos(\pi k_{3})$%
\end{tabular}
\\\hline
$DM_{[4,6]}$ &
\begin{tabular}
[c]{ccc}%
correct &  & $-4D_{2}\sin(\pi k_{1})\sin(\pi k_{3})-4iE_{2}\sin(\pi k_{2}%
)\cos(\pi k_{1})+4iE_{2}\sin(\pi k_{2})\cos(\pi k_{3})$\\
wrong &  & $-4D_{2}\sin(\pi k_{1})\sin(\pi k_{3})+4iE_{2}\sin(\pi k_{1}%
)\cos(\pi k_{2})-4iE_{2}\sin(\pi k_{3})\cos(\pi k_{2})$%
\end{tabular}
\\\hline
$DM_{[5,6]}$ &
\begin{tabular}
[c]{ccc}%
correct &  & $-4D_{2}\sin(\pi k_{2})\sin(\pi k_{3})-4iE_{2}\sin(\pi k_{1}%
)\cos(\pi k_{2})+4iE_{2}\sin(\pi k_{1})\cos(\pi k_{3})$\\
wrong &  & $-4D_{2}\sin(\pi k_{2})\sin(\pi k_{3})+4iE_{2}\sin(\pi k_{2}%
)\cos(\pi k_{1})-4iE_{2}\sin(\pi k3)\cos(\pi k1)$%
\end{tabular}
\\\hline\hline
\end{tabular}
\end{table}


\begin{thebibliography}{99}                                                                                               %


\bibitem {PRB 47 1397 1993}B. Delley and E.F. Steigmeier, Phys. Rev. B 47,
1397 (1993).

\bibitem {SCIENCE 262 1242 1993}William L. Wilson, P. F. Szajowski, and L. E.
Brus, SCIENCE 262, 1242 (1993).

\bibitem {J Phys Chem 97 1224 1993}K. A. Littau, P. J. Szajowski, A. J.
Muller, A. R. Kortan, and L. E. Brus, J. Phys. Chem. 97, 1224 (1993).

\bibitem {JAP 75 4486 1994}R. F. Pinizzotto, H. Yang, J. M. Perez and J. L.
Coffer, J. App. Phys. 75, 4486 (1994).

\bibitem {PRL 81 2803 1998}D. Kovalev, H. Heckler, M. Ben-Chorin, G. Polisski,
M. Schwartzkopff, and F. Koch, Phys. Rev. Lett. 81, 2803 (1998).

\bibitem {JAP 105 023108 2009}P. Bianucci, J. R. Rodr\'{\i}uez, C. M.
Clements, J. G. C. Veinot and A. Meldrum, J. App. Phys. 105, 023108 (2009).

\bibitem {laser photonic review}Nicola Daldosso and Lorenzo Pavesi, Laser \&
Photon. Rev. 3, 508 (2009).

\bibitem {J Lum 80 263 1999}Nenad Lalic and Jan Linnros, J. Luminescence 80,
263 (1999).

\bibitem {Superlattices and Microstructures 28 177 2000}Y. Fu, M. Willander,
A. Dutta, and S. Oda, Superlattices and Microstructures 28, 177 (2000).

\bibitem {JAP 78 6705 1995}Hua Xia, Y. L. He, L. C. Wang, W. Zhang, X. N. Liu,
X. K. Zhang, and D. Feng, J. Appl. Phys. 78, 6705 (1995).

\bibitem {J Phys Condens Matter 13 L835 2001}Xinhua Hu, Guozhong Wang, Weimin
Wu, Ping Jiang and Jian Zi, J. Phys. Condens. Matter 13, L835 (2001).

\bibitem {J Phys Condens Matter 14 L671 2002}Xinhua Hu and Jian Zi, J. Phys.:
Condens. Matter 14, L671 (2002).

\bibitem {J Phys Condens Matter 20 145213 2008}Audrey Valentin, Johann
S\'{e}e, Sylvie Galdin-Retailleau and Philippe Dollfus, J. Phys.: Condens.
Matter 20, 145213 (2008).

\bibitem {PRB 80 193410 2009}Giuseppe Faraci, Santo Gibilisco, and Agata R.
Pennisi, Phys. Rev. B 80, 193410 (2009).

\bibitem {nanotech 18 175705 2007}P Roura, J Farjas, A Pinyol, and E Bertran,
Nanotechnology 18, 175705 (2007).

\bibitem {JAP 90 4175 2001}G. Viera, S. Huet, and L. Boufendi, J. App. Phys.
90, 4175 (2001).

\bibitem {Chem Phys Letter 382 502 2003}F.M. Liu, B. Ren, J.H. Wu, J.W. Yan,
X.F. Xue, B.W. Mao, and Z.Q. Tian, Chem. Phys. Lett. 382, 502 (2003).

\bibitem {PRB 64 073304}Puspashree Mishra and K.P. Jain, Phys. Rev. B 64,
073304 (2001).

\bibitem {Richter's}H.Richter, Z.P. Wang, and L.Ley, Solid State Commun. 39,
625 (1981).

\bibitem {SSC Campbell and Fauchet}I.H. Campbell and P.M. Fauchet, Solid State
Commun. 58, 739 (1986).

\bibitem {JAP 86 1921 1999}V. Paillard, P. Puech, M.A. Laguna, R. Carles, B.
Kohn, and F. Huisken, J. App. Phys. 86, 1921 (1999).

\bibitem {PRB 73 033307 2006}Giuseppe Faraci, Santo Gibilisco, Paola Russo,
and Agata R. Pennisi, Phys. Rev. B 73, 033307 (2006).

\bibitem {BPA Bell}R. J. Bell and D. C. Hibbins-Butler, J. Phys. C: Solid
State Phys. 9, 2955 (1976).

\bibitem {Cheng and Ren}Wei Cheng and Shang-Fen Ren, Phys. Rev. B 65, 205305 (2001).

\bibitem {BPA fullerene}S. Guha, J. Men\'{e}ndez, J.B. Page, and G.B. Adams,
Phys. Rev. B 53, 13106 (1996).

\bibitem {PRB 66 161311R 2002 local heating}M. J. Konstantinovi\'{c}, S.
Bersier, X. Wang, M. Hayne, P. Lievens, R. E. Silverans, and V. V.
Moshchalkov, Phys. Rev. B 66, 161311(R) 2002.

\bibitem {RIM}K. Kunc, M. Balkanski, and M. A. Nusimovici, Phys. Stat. Sol.
(b) 71, 341 (1975).

\bibitem {book methods in computational physics}G. Dolling, in \textit{Methods
in Computational Physics}, edited by Berni Alder, Sidney Fernbach, and Manuel
Rotenberg, (Academic, New York, 1976), Vol. 15, p. 26.

\bibitem {PRB 74 054302 2006}M. Aouissi, I. Hamdi, N. Meskini and A. Qteish,
Phys. Rev. B 74, 054302 (2006).

\bibitem {Phys Rev B 55 10337 1997}Xavier Gonze, Phys. Rev. B 55, 10337 (1997).

\bibitem {Phys Rev B 55 10355 1997}Xavier Gonze and Changyol Lee, Phys. Rev. B
55, 10355 (1997).

\bibitem {Computer Phys Commun 180 2582 2009}Xavier Gonze, et al., Computer
Phys. Commun. 180, 2582 (2009).

\bibitem {Zeit Kristallogr 220 558 2005}Xavier Gonze et al., Zeit.
Kristallogr. 220, 558 (2005).

\bibitem {data1}J. Kulda, D. Strauch, P. Pavone and Y. Ishii, Phys. Rev. B 50,
13347 (1994).

\bibitem {data2}G. Nilsson and G. Nelin, Phys. Rev. B 6, 3777 (1972).

\bibitem {even grided k points}Hendrik J. Monkhorst and James D. Pack, Phys.
Rev. B 13, 5188 (1976).

\bibitem {method of relaxation}Raju P. Gupta, Phys. Rev. B 23, 6265 (1981).

\bibitem {book D A LONG}Derek A. Long, in \textit{The Raman Effect, A Unified
Treatment of the Theory of Raman Scattering by Molecules,} p. 101 (Wiley,
Chichester, 2002).

\bibitem {SY lin paper 1}Y. C. Liao, S. Y. Lin, S. C. Lee and C. T. Chia,
Appl. Phys. Lett. 77, 4328 (2000).

\bibitem {SY lin paper 2}C. W. Lin, S. Y. Lin, S. C. Lee and C. T. Chia, J.
Appl. Phys. 91, 1525 (2002).

\bibitem {SY lin paper 3}C. W. Lin, S. Y. Lin, S. C. Lee and C. T. Chia, J.
Appl. Phys. 91, 2322 (2002).

\bibitem {TED and TEM graphics}Shu-Ting Chou et. al., \textquotedblleft
Structural and Optical Properties of Silicon Nanoparticles Prepared by Thermal
Evaporation\textquotedblright, (conference) OPT 2007, Taichung, Taiwan,
Republic of China.

\bibitem {group theory Inui}T. Inui, Y., Tanabe, and Y. Onodera, in
\textit{Group Theory and Its Applications in Physics }(Springer, Berlin, 1990).

\bibitem {two phonons in si bulk}Paul A. Temple and C.E. Hathaway, Phys. Rev.
B 7, 3685 (1973).
\end{thebibliography}

\begin{thebibliography}{9999}                                                                                             %


\bibitem {chang wrong paper 1}S.-F. Ren, H. Chu, and Y. C. Chang, Phys. Rev.
Lett. 59, 1841 (1987).

\bibitem {chang wrong paper 2}S.-F. Ren, H. Chu, and Y.-C. Chang, Phys. Rev.
B, 37, 8899 (1988).

\bibitem {chang wrong paper 3}H. Chu, S.-F. Ren, and Y. C. Chang, Phys. Rev. B
37, 10746 (1988).

\bibitem {kunc RIM CPC}K. Kunc and O. Holm Nielsen, Computer Phys. Commu. 16,
181 (1979).

\bibitem {kunc previous}K. Kunc, M. Balkanski, and M. A. Nusimovici, Phys.
Stat. Sol. (b) 71, 341 (1975).

\bibitem {J Phys Conden Matt 2 1457 1990}D. Strauch and B. Dorner, J. Phys.:
Condens. Matter 2, 1457 (1990).
\end{thebibliography}
\end{document}